\documentclass[12pt,a4paper]{article}

\newcommand{\be}{\begin{equation}}
\newcommand{\ee}{\end{equation}}
\newcommand{\bea}{\begin{eqnarray}}
\newcommand{\eea}{\end{eqnarray}}

\newcommand{\Tr}{\mathop{\mathrm{Tr}}\nolimits}
\newcommand{\dd}{\mathrm{d}}

\begin{document}

\title{Canonical Quantization of Noncommutative Field Theory}

\author{Ciprian Acatrinei\thanks{On leave from: {\it Institute of
        Atomic Physics  -
        P.O. Box MG-6, 76900 Bucharest, Romania}; e-mail:
        acatrine@physics.uoc.gr.} \\
        Department of Physics, University of Crete, \\
        P.O. Box 2208, Heraklion, Greece}

\date{April 23, 2002}

\maketitle

\begin{abstract}
A simple method to canonically quantize noncommutative field theories is proposed.
As a result, 
the elementary excitations of a $(2n+1)$-dimensional scalar field theory
are shown to be 
bilocal objects
living in an $(n+1)$-dimensional space-time. 
Feynman rules for their scattering are derived canonically.
They agree, upon suitable redefinitions, with the rules obtained
via star-product methods.
The IR/UV connection is interpreted within this framework.
\end{abstract}

{\bf Introduction and Summary}

Noncommutative field theories  \cite{reviews} are interesting, 
nonlocal but 
most 
probably consistent, extensions of the usual ones.
They also arise as a particular low energy limit of string theory \cite{st1,st2}.
The fields are defined over a  base space which is noncommutative
\cite{reviews}, 
often obeying relations of the type 
$[x_{\mu},x_{\nu}]=i\theta_{\mu\nu}$.
At the classical level,  new physical features appear
in these theories. For instance,
one encounters solitonic excitations in higher dimensions
\cite{solitons}, superluminal propagation \cite{speed_of_light}, 
or waves propagating on discrete spaces \cite{radial_waves}.
At the quantum level, one has two superimposed structures:
the coordinate space, where  $[\hat{x}_{\mu},\hat{x}_{\nu}]\neq 0$,
and the dynamical fields' (fiber) space, where  canonically conjugate
variables  do not commute, 
$[\hat{\phi}(t,\vec{x}),\hat{\pi}(t,\vec{x}')]\neq 0$.	

This two-level structure hampered the canonical quantization of
noncommutative (NC) field theories.  
Consequently, their perturbative quantum dynamics has been
studied via star-product techniques \cite{reviews},
i.e. by replacing operator products with the Groenewold-Moyal one.
This leads to deformed theories, living on a commutative space of Weyl symbols. 
Perturbation theory is then defined in the usual way.
Loop calculations performed in this set-up pointed to 
an intriguing mixing between short distance and long distance physics, 
called the IR/UV connection \cite{ir_uv,ir_uv_2,ir_uv_3}. 

The purpose of this paper is to develop simple canonical techniques
for the direct quantization of noncommutative fields. 
We present here the basic idea, describe the nature of the degrees of freedom
and their rules of interaction, as well as some implications.
Our motivations are at least two-fold.
First, phase space quantization  methods \cite{zachos}
are not always the most useful ones, either for particles or for fields.
Actually, commutative quantum theories developed 
mostly through canonical, functional, or propagator methods.
Second, 
canonical quantization offers a clear picture of the degrees of freedom of a theory, 
picture which is not rigorously established in NC spaces,
in spite of many interesting works \cite{dipoles, dipoles_2}.
Our elementary operatorial methods will automatically
lead to such a picture. We show that
the fundamental excitations of a $(2n+1)-$dimensional scalar theory (with commuting time)
are bilocal objects living in a lower, $(n+1)-$, dimensional space-time.
We will call them rods, or dipoles, 
although no charge of any kind enters their description. 
The information on the remaining $n$ spatial directions is encoded into the
length and orientation of the dipoles. Those $n$ parameters are, in turn, proportional
to the momentum a noncommutative particle would have in the `lost' directions.
This picture puts on a firmer ground a general belief 
\cite{dipoles, dipoles_2} 
that noncommutative theories are about dipoles, not particles.
Moreover, it shows that these dipoles live in a lower dimensional space. 
Rules for their propagation and scattering are obtained canonically. 
They show that the above 
dimensional reduction is limited
to tree level dynamics: the loop integrations are taken also over
the dipole parameters, restoring the $(2n+1)$-dimensionality of the theory,
as far as renormalization is concerned. 
Upon identification of the rod parameters with the momenta in the conjugate directions, 
our Feynman rules agree
with the ones obtained a long time ago through star product technology.
The physical interpretation is however different, 
being hopefully more intuitive and adequate for the description of experiments. 
The  interpretation of the IR/UV mixing given in \cite{ir_uv_2}
can be adapted to this framework.
One also notices that  interaction `vertices' for dipoles
have in general finite area, and a poligonal boundary.
As far as this area is kept finite,  loop amplitudes are effectively regulated
by noncommutativity. 
However, once this area shrinks to zero 
(in planar diagrams, or nonplanar ones with zero external momentum),
the NC phase is of no effect, and UV infinities are present.
They metamorphose into IR divergences if the cause of the vertex shrinking
is an external momentum going to zero.

{\bf Bilocal objects}

Let us consider a $(2+1)$-dimensional scalar field 
$\Phi(t,\hat{x},\hat{y})$ defined over a commutative time $t$ and 
a pair of NC coordinates satisfying
\be
[\hat{x},\hat{y}]=i\theta. \label{nc}
\ee
The extension to $n$ NC pairs is straightforward.
Commutative spatial directions are dropped, for simplicity.  
The action is
\be
S=\frac{1}{2}\int\dd t \Tr_{{\cal H}} 
\left [
\dot{\Phi}^2
-(\partial_x \Phi)^2 -(\partial_y \Phi)^2
-m^2\Phi^2- 2 V(\Phi)
\right ].
\label{action}
\ee    
$\hat{x}$ and $\hat{y}$ act on a harmonic oscillator Hilbert space ${\cal H}$  
in the usual way. ${\cal H}$ may be given a discrete basis $\{|n>\}$ 
formed by eigenstates of $\hat{x}^2+\hat{y}^2$,
or a continuous one $\{|x>\}$, 
composed of eigenstates of, say, $\hat{x}$. 
In what follows, we will discuss explicitely quartic potentials, 
$V(\Phi)=\frac{g}{4!}\Phi^4$. 
Cubic potentials are actually simpler, but maybe less relevant physically.

We (`doubly'-)quantize the field $\Phi$ by writing
\be
\Phi=\int\int\frac{dk_x dk_y}{2\pi\sqrt{2\omega_{\vec{k}}}}
\left [
\hat{a}_{k_x k_y}e^{i(\omega_{\vec{k}}t-k_x\hat{x}-k_y\hat{y})}
+\hat{a}^{\dagger}_{k_x k_y}e^{-i(\omega_{\vec{k}}t-k_x\hat{x}-k_y\hat{y})}
\right ]. 
\label{Qf}
\ee
$\hat{x}$ and $\hat{y}$ are operators acting on the Hilbert space ${\cal H}$,
which appeared  due to their noncommutativity. 
$\hat{a}_{k_x k_y}$ and $\hat{a}^{\dagger}_{k_x k_y}$
act on the usual 
Fock space  ${\cal F}$ of a quantum field theory (FT).
We have thus a `doubly'-quantum FT,
with $\Phi$ acting on a direct product of two Hilbert spaces,
namely ${\cal F}\otimes {\cal H}$. 
To prove (\ref{Qf}),  start with a classical field living
on a commuting space. Upon usual field quantization, $a$ and $a^{*}$
become operators on the Fock space ${\cal F}$.
To make the underlying space noncommutative, introduce (\ref{nc})
and apply the Weyl quantization procedure \cite{weyl}
to the exponentials  $e^{i(k_x x+k_y y)}$. The result is (\ref{Qf}), which
means the following: $\Phi$ creates (destroys), 
via $\hat{a}^{\dagger}_{k_xk_y}$ ($\hat{a}_{k_xk_y}$), an excitation 
represented by a "plane wave" 
$e^{i(\omega_{\vec{k}}t-k_x\hat{x}-k_y\hat{y})}$.
We will now describe such an object. 

We could work with $\Phi$ as an operator ready to act on both Hilbert spaces 
${\cal F}$ and ${\cal H}$ .
It is however simpler to "saturate" it on ${\cal H}$, working with expectation values
$<x'|\Phi|x>$, which can still act on ${\cal F}$. 
$|x>$ is an eigenstate of $\hat{x}$, $\hat{x}|x>=x|x>$,
$\hat{y}|x>=-i\theta\frac{\partial}{\partial x}|x>$.
This means keeping only one coordinate out of a pair of NC spatial directions
(for $n$ pairs, commutativity is gained on the reduced space at the expense of strict locality).
A key equation is now
\be
<x'|e^{i(k_x\hat{x}+k_y\hat{y}}|x>
=e^{ik_x(x+k_y\theta/2)}\delta(x'-x-k_y\theta)
=e^{ik_x\frac{x+x'}{2}}\delta(x'-x-k_y\theta). \label{bilocal}
\ee
This is a bilocal expression, 
and we already see that its span along the $x$ axis,
$(x'-x)$, is proportional to the momentum along  the conjugate $y$ direction, 
i.e. $(x'-x)=\theta k_y$.
Using (\ref{Qf},\ref{bilocal}), one sees that 
\be
<x'|\Phi|x>=\int \frac{dk_x}{2\pi\sqrt{2\omega_{k_x,k_y}}}
\left [
\hat{a}_{k_x, k_y}e^{i(\omega_{\vec{k}}t-k_x\frac{x+x'}{2})}
+\hat{a}^{\dagger}_{k_x ,-k_y}e^{-i(\omega_{\vec{k}}t+k_x\frac{x+x'}{2}}
\right ].
\label{bif}
\ee
where $k_y=(x'-x)/\theta$.
Thus, $\Phi$ annihilates  a rod of momentum $k_x$ and length $\theta k_y$,
and creates a rod of momentum $k_x$ and length $-\theta k_y$.
Due to (\ref{nc}), one degree of freedom apparently disappears from (\ref{bif}).
Its presence shows up only through the modified dispersion relation 
\be
\omega_{(k_x, k_y=\frac{x'-x}{\theta})}=\sqrt{k_x^2+\frac{(x'-x)^2}{\theta^2}+m^2}.\label{energy}
\ee

{\bf Correlators}

Let us now calculate two-point correlation functions for such rods. 
The expectation value of the product of two bilocal fields,
taken on the Fock space ${\cal F}$ vacuum  $|0\rangle$, is
\be
\langle 0| <x_4|\Phi|x_3> <x_2|\Phi|x_1> |0\rangle =
\int \frac{dk_x}{8\pi^2 \omega_{k_x,}}
e^{ik_x[\frac{x_3+x_4}{2}-\frac{x_1+x_2}{2}]}\delta(x_4-x_3-x_2+x_1)
\label{propagator}
\ee
where $k_y=(x'-x)/\theta$, and 
$\omega_{k_x,k_y=(x'-x)/\theta}$ obeys (\ref{energy}) again.
Again, there is no integral along $k_y$. More precisely,
if one compares (\ref{propagator}) to the $(1+1)$-dimensional commutative
correlator of two fields,
$\langle0|\phi(X_2)\phi(X_1)|0\rangle$, with $X_1=(x_1+x_2)/2$ and  $X_2=(x_3+x_4)/2$,
the only differences are
the additional $\frac{(x'-x)^2}{\theta^2}$ term
in (\ref{energy}), and the delta function $\delta([x_4-x_3]-[x_2-x_1])$,
which ensures that the length of the rod (the momentum along $y$) is conserved.
Thus, our bilocal objects propagate in a $(1+1)-$dimensional space. 
The extra $y$ direction
is accounted for by their lenght - which contributes to the energy, 
and orientation. We will also call these rods dipoles,
although they have no charges at their ends
(at least for real scalar fields), 
and they are extended objects in the absence of any background.
One may
speculate on possible relations of these rods with stretched open strings,
or with the double index representation for Yang-Mills theories.

{\bf Interactions}

The quartic interaction term in (\ref{action}) can be written as
\be
\int dt Tr_{{\cal{H}}} V(\Phi)
=\frac{g}{4!}\int dt \int_{x,a,b,c}
<x|\Phi|a><a|\Phi|b><b|\Phi|c><c|\Phi|x>.
\label{potential}
\ee
We will have a look at some terms in the Dyson series 
generated by (\ref{potential}),
to illustrate the canonical derivation of the Feynman rules. 
Let $:\hat{A}\hat{B} :$ denote normal ordering  of $\hat{A}\hat{B}$.
Once the vacuum correlator (\ref{propagator}) is known,
the derivation of the diagrammatic rules follows the
standard procedure; hence we will not
present it in detail.
To find the basic `vertex' for four-dipole scattering we evaluate
\be
\langle -\vec{k}_3, -\vec{k}_4|
:\int dt \int_{x,a,b,c}
<x|\Phi|a><a|\Phi|b><b|\Phi|c><c|\Phi|x> :
|\vec{k}_1, \vec{k}_2 \rangle
\label{vertex}
\ee
$|\vec{k}_1, \vec{k}_2 \rangle$ is a Fock space state, meaning two quanta are present, 
with momenta $\vec{k}_1$ and $\vec{k}_2$.
The momenta $\vec{k}_{i, i=1,2,3,4}$ have each two components: $\vec{k}_i=(k_i,l_i)$.
$k_i$ is the momentum along $x$, whereas $l_i$ represents the dipole
extension along $x$ (corresponding to the momentum along $y$) . 
Using Eq. (\ref{bif}) and integrating over $x,y,z$ and $u$,
one obtains the conservation laws $k_1+k_2=k_3+k_4$ and $l_1+l_2=l_3+l_4$.
The final result differs from the  four-point scattering
vertex of $(2+1)$ commutative particles with  momenta $\vec{k}_i=(k_i,l_i)$
only through the phase
\be
e^{-\frac{i\theta}{2}\sum_{i<j}(k_i l_j-l_i k_j)}. \label{phase}
\ee
Interpreting $l_i$ as the $i$-th momentum along $y$, 
this is precisely the star-product modification of the usual Feynman rules.
The phase (\ref{phase}) appears due to the bilocal nature
of generic $<x'|\Phi|x>$'s. Pointlike $<x|\Phi|x>$'s would never produce it.

By contracting adjacent (nonadjacent) terms in (\ref{vertex}), one obtains the 
planar (nonplanar) one-loop correction to the free rod propagator,
together with the recipe for calculating loops. Again, the derivation is 
straightforward.
The main result is that
one has to integrate over both the momentum and length of the dipole circulating
in a loop. This $\frac{1}{2\pi}\int dk_{loop} \int dl_{loop}$ integration,
together with the dispersion relation (\ref{energy}),
brings back into play - as far as divergences are concerned - the $y$
direction.
It is easy to extend the above reasoning to $(2n+1)-$dimensions:
unconstrained  dipoles will propagate in a $(n+1)$-dimensional commutative space-time;
their Feynman rules are obtained as outlined above. 
Once the dipole lengths are interpreted as momenta in the conjugate directions, 
our rules are identical to those obtained long ago  via star-product calculus.
The calculational aspects have been extensively explored 
\cite{reviews, ir_uv,ir_uv_2, ir_uv_3} in the last years.
Our interpretation is however different, and in this light, we will discuss now the 
IR/UV connection.  

{\bf IR/UV}

We have derived directly from the field theory the dipolar character of
the NC scalar field excitations.
We saw that, in the $\{|x>\}$ basis, the momentum in the conjugate direction
becomes the lenght of the dipole.
Thus, a connection between ultraviolet (large momentum) and infrared
physics (large distances) becomes evident. This puts on a more rigorous basis the 
argument of \cite{ir_uv_2} concerning the IR/UV connection.

Moreover, we can provide a geometrical view of 
the differences between planar and nonplanar loop diagrams,
and the role of low momenta in nonplanar graphs.
Let us go to $(4+1)$ directions, $t,\hat{x},\hat{y},\hat{z},\hat{u}$, and assume
$[\hat{x},\hat{y}]=[\hat{z},\hat{w}]=i\theta$. Consider a $\{|x,z>\}$ basis.
Then we can speak of a commutative space spanned by the axes $x$ and $z$, 
on which dipoles with momentum $\vec{p}=(p_x,p_z)$ and length
$\vec{l}=(l_x,l_z)=\theta(p_y,p_w)$ evolve. Consider the scattering of four such dipoles,
Their `meeting place' is  a poligon with four edges and area ${\cal A}$
(figure 1a).

\begin{picture}(350,230)(10,-20)
\thicklines
\put(40,110){\vector(1,1){40}}
\put(80,150){\vector(1,-1){50}}
\put(130,100){\vector(-2,-1){74}}
\put(56,63){\vector(-1,3){16}}
\put(32,173){\vector(2,-3){15}}
\put(120,135){\vector(4,3){20}}
\put(100,76){\vector(4,-1){20}}
\put(30,35){\vector(1,3){12}}
\put(230,140){\vector(-1,2){20}}
\put(212,182){\vector(1,-2){20}}
\put(250,130){\vector(-2,1){20}}
\put(232,142){\vector(2,-1){20}}
\put(220,85){\vector(2,1){30}}
\put(250,100){\vector(1,-4){10}}
\put(260,60){\vector(-2,-1){30}}
\put(230,45){\vector(-1,4){10}}
\put(320,90){\vector(2,1){4}}
\put(324,92){\vector(1,-4){10}}
\put(334,52){\vector(-2,-1){4}}
\put(330,50){\vector(-1,4){10}}
\thinlines
\put(40,110){\line(-2,3){20}}
\put(80,150){\line(-2,3){20}}
\put(80,150){\line(4,3){30}}
\put(130,100){\line(4,3){30}}
\put(130,100){\line(4,-1){30}}
\put(56,63){\line(4,-1){30}}
\put(56,63){\line(-1,-3){15}}
\put(40,110){\line(-1,-3){15}}
\put(67,105){${\cal A}\neq 0$}
\put(225,65){${\cal A}\neq 0$}
\put(250,160){${\cal A}= 0$}
\put(340,70){${\cal A}= 0$}
\put(150,140){\vector(1,0){50}}
\put(155,150){planar}
\put(152,55){nonplanar}
\put(150,70){\vector(1,0){50}}
\put(270,70){\vector(1,0){40}}
\put(270,55){$\vec{l}_{ext}\rightarrow 0$}
\put(221,165){$\vec{l}_{ext}$}
\put(220,100){$\vec{l}_{ext}$}
\put(240,140){$\vec{l}_{loop}$}
\put(256,88){$\vec{l}_{loop}$}
\put(42,122){$\vec{l}_{1}$}
\put(49,89){$\vec{l}_{2}$}
\put(70,78){$\vec{l}_{3}$}
\put(94,139){$\vec{l}_{4}$}
\put(39,169){$\vec{k}_{1}$}
\put(20,40){$\vec{k}_{2}$}
\put(94,59){$-\vec{k}_{3}$}
\put(119,152){$-\vec{k}_{4}$}
\put(80,30){(a)}
\put(240,30){(c)}
\put(225,186){(b)}
\put(330,30){(d)}
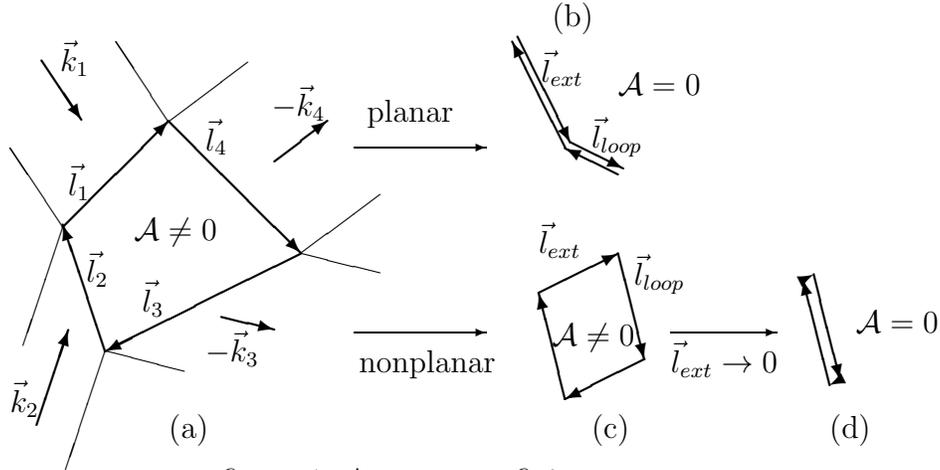
\put(100,10){figure 1: Area versus finiteness}
\end{picture}

\noindent
One has two possibilities for the one-loop correction to the propagator:
planar and nonplanar.
In the planar case, adjacent dipole fields are contracted. Momentum and
length conservation enforce then the poligon to degenerate into a
one-dimensional, zero-area object (figure 1b). UV divergences persist.
In the nonplanar case, due to the nonadjacent contraction
the area ${\cal A}$ does not go to zero (cf. figure 1c)
unless the external dipole length vanishes (figure 1d). 
${\cal A}\neq 0$ appears thus to be related to the disappearance of UV divergences.
Actually, the true regulator is the phase (\ref{phase}).
This is zero, i.e. ineffective, when ${\cal A}= 0$
in {\it both} the $|x,z>$ and $|y,u>$ bases.
That corresponds to zero external length {\it and} momentum
in the dipole picture, which means that the resulting divergence is 
half IR ($\vec{p}_{ext}=0$) and half UV ($\vec{l}_{ext}=0$)!
In Weyl space this is just the usual zero external momentum,
say $p_{\mu}^{ext}=0$ - interpreted there as an IR divergence.
For dipoles the divergence comes from having zero vertex area ${\cal A}$ in any basis,
and is half IR and half UV.
NCFT is somehow between usual FT and string theory:
when the interaction vertex is a point, UV infinities appear;
when it opens up, as in string theory, amplitudes are finite. 

{\bf Remarks}


We saw that by dropping $n$ coordinates, intuition is gained:
the remaining space admits a notion of distance,
although bilocal (and in some sense IR/UV dual) objects probe it.
Other bases of ${\cal{H}}$ can also be used. 
For instance, the basis $\{|n>\}$, formed by eigenvectors of  $\hat{n}=\sqrt{x^2+y^2}$,
leads to a discrete remnant space \cite{radial_waves}.
Although the phase operator conjugated to $\hat{n}$ is not easy to define,
the multilocal character of the excitations is preserved. 

One could put the scalar fields on a torus by imposing periodic boundary conditions. 
In this case, (discrete) high momenta along $y$ would correspond to dipoles which wind
around the circle spanned by $x$. This relationship between winding and momentum states
is reminescent of T-duality, and suggests that the canonical description may be employed 
in describing Morita equivalence.

An important  question is: how do the dimensionality and noncommutativity of space-time
depend on the regime in which we probe the theory?
To start, we have a NC $(2n+1)-$dimensional theory. Then, at tree level
(i.e. classical plus tree level interference effects), 
one has $D=n+1$ commuting directions. 
However, loop effects drive us back to $D=2n+1$.
At a scale $r\sim \sqrt{\theta}$, space is surely NC. 
For  $r>> \sqrt{\theta}$ it is believed to be commutative. 
However,
if $r$ is the radius in the largest available commutative subspace,
the IR/UV connection 
suggests a connection (duality?) between the $r>> \sqrt{\theta}$ and $r<< \sqrt{\theta}$ regimes.
A clarification of these issues is desirable.

\vskip 0.3cm

In conclusion, we found a simple way to quantize scalar NCFT through canonical methods.
This provides a quantitative description for the kinematics and dynamics of such theories
- including  limits in the dimensional reduction one may hope for, 
and a simple reinterpretation of the IR/UV connection.
Although the Feynman rules derived in this way were previously known and used,
we believe we provided a simple and clear picture for the degrees of freedom of the theory.
This alternative point of view may find interesting applications, e.g. along the lines
sketched in the above remarks.
An extension of the method to gauge theories, as well as a path integral approach,
are presently under study.

\vskip 0.3cm

{\bf Acknowledgments} 

I benefited from a pleasant
working environment within the HEP Theory Group of the University of Crete.
This work was supported through a European Community Marie Curie fellowship,
under Contract HPMF-CT-2000-1060.


\end{document}